\documentclass[nofootinbib,notitlepage,superscriptaddress,10pt,aps,prd,onecolumn]{revtex4-1}
\usepackage[utf8]{inputenc}
\usepackage{amsmath,mathtools}

\usepackage{tensor}
\usepackage{amssymb}
\usepackage{graphicx}
\usepackage{physics}
\usepackage{hyperref}
\usepackage{natbib}
\usepackage{lipsum}  
\usepackage{bigints}
\usepackage{verbatim}
\newcommand{\leri}[1]{\left(#1\right)}

\begin{document}
\title{Gravitational Landau Damping for massive scalar modes}
\author{Fabio Moretti}
\email{fabio.moretti@uniroma1.it}
\affiliation{Physics Department, ``Sapienza'' University of Rome, P.le Aldo Moro 5, 00185 (Roma), Italy}
\author{Flavio Bombacigno}
\email{flavio.bombacigno@ext.uv.es}
\affiliation{Departament de F\'{i}sica Teòrica and IFIC, Centro Mixto Universitat de València - CSIC, Universitat de València, Burjassot 46100, València, Spain}
\author{Giovanni Montani}
\email{giovanni.montani@enea.it}
\affiliation{Physics Department, ``Sapienza'' University of Rome, P.le Aldo Moro 5, 00185 (Roma), Italy}
\affiliation{ENEA, Fusion and Nuclear Safety Department, C. R. Frascati,
	Via E. Fermi 45, 00044 Frascati (Roma), Italy}

\begin{abstract}
We establish the { possibility of  Landau damping for gravitational scalar waves which propagate in a non-collisional gas of particles. In particular, under the hypothesis of homogeneity and isotropy, we describe the medium at the equilibrium with a J\"uttner-Maxwell distribution, and we analytically determine the damping rate from the Vlasov equation. We find that damping occurs only if the phase velocity of the wave is subluminal throughout the propagation within the medium. Finally, we investigate relativistic media in cosmological settings by adopting numerical techniques.}
\end{abstract}
	\maketitle
\section{Introduction}
\noindent One of the most surprising features of a plasma, considered as a dielectric medium, is the attenuation of electromagnetic waves even when collisions can be neglected. This phenomenon, known as the "Landau damping'' \cite{Landau:1946jc}, is essentially due to the presence of a long range interaction, which affects a large number of particles contained in the so called Debye sphere \cite{Debye_1923}. This situation resembles to some extent the interaction of gravitational waves with a material medium, despite two basic properties are here missing with respect to the electromagnetic counterpart. In particular, we refer to the neutralization of the background, which for a plasma is provided by the ion distribution, and to longitudinal excitations, which appear by virtue of the effective mass acquired by photons when crossing a plasma \cite{PhysRevE.49.3520,Mendonca:2000tk}. The first of these discrepancies can be locally overcome, since the role of the neutralizing background can be played by inertial forces. Nearby a spacetime event, indeed, Christoffel symbols associated to the background, which enter the Vlasov equation, can be made almost vanishing by choosing a local inertial frame. In the linear theory considered below, this fundamental point of view is { enclosed} by applying the model to a region of space where homogeneity and isotropy of the medium can be assumed (see \cite{1962MNRAS.124..279L,1971SvA....14..758B,1973SvA....16..830M} for pioneering treatments). By other words, we { only consider gravitational perturbations far smaller than the characteristic spatial scale of the background}. Concerning the absence of longitudinal modes for gravitational waves, instead, we observe that when gravitational subsystems are properly treated as molecular media \cite{Szekeres:1971ss,Montani:2018iqd} or following a hydrodynamic and kinetic approach \cite{Anile,Ehlers:1987nm,Prasanna:1999pn,Barta:2017vip,1978SvA....22..528Z}, additional polarizations can be typically excited. Although this effect is conceptually very relevant, it is in reality very small and associated to peculiar wavelengths. More intriguing, therefore, is to look at the large number of modified theories of gravity which allow for the emergence of an additional massive scalar { mode}, as for instance { Horndeski} gravity \cite{1974IJTP...10..363H,2011PThPh.126..511K,2009PhRvD..79h4003D,Hou:2017bqj}, hybrid metric-Palatini approaches \cite{PhysRevD.85.084016,Capozziello:2013uya,Capozziello:2012qt,Capozziello:2013yha,PhysRevD.87.084031,PhysRevD.95.124035,Bombacigno:2019did} or massive gravity \cite{1939RSPSA.173..211F,2010PhRvD..82d4020D,2011PhRvL.106w1101D,2012JHEP...02..026H}. In this case, in fact, together with a breathing polarization in the transverse plane, scalar fields are responsible for a longitudinal polarization as well, that we expect could interact with particles of the medium.
\\ Now, it is a well established result that transverse gravitational waves are not absorbed by non-collisional massive media \cite{osti_4641583,Gayer:1979ff,PhysRevD.13.2724,Flauger:2017ged}. { Damping is indeed only possible in the presence of viscosity \cite{Hawking:1966qi,Madore1,Anile}, or when a medium of massless particles is considered \cite{Chesters:1973wan,Weinberg:2003ur,PhysRevD.88.083536}.} Then, having in mind that { the presence of} longitudinal modes could imply the gravitational analog of Landau damping, we { adopt} a kinetic theory approach { and we analyze} the interaction of scalar waves with a { collisionless} particle distribution. { Then, we calculate} the gravitational Langmuir dispersion relation, { denoting} self-consisting fluctuations in the medium, { and we determine the damping from} the imaginary part of the frequency. { In this respect, we show that the damped scalar mode naturally decouples from the non-damped transverse tensor polarizations. Furthermore, we find that the phase velocity of the scalar mode must} be subluminal, otherwise the typical poles inducing the Landau damping fall out the allowed domain. In particular, this condition reflects in a specific phenomenological inequality, relating the traversed medium with parameters describing the model taken into account. { Especially,} we show that this is just the reason for which standard transverse gravitational wave are not absorbed, violating the condition above.
\\ { The paper is structured as follows. In \autoref{sec1} we introduce the theoretical framework and the fundamental equations describing the propagation of the gravitational modes. In \autoref{sec2} we derive, starting from the Vlasov equation, the damping rate, showing as the tensor modes decouple from the scalar excitation. In \autoref{sec3} we implement a numerical investigation for relativistic media, and we explicitly evaluate the damping for a cosmological scenario. Finally, considerations are drawn in \autoref{sec4}.}
\section{Decoupling of gravitational modes}\label{sec1}
 We are interested in wave propagation for extended theories of gravity, where the settling of a scalar mode is predicted besides standard tensor excitations. It seems reasonable, therefore, to look at the so called Horndeski models, which represent the most general class of scalar tensor theories endowed with higher order derivative terms in the action which preserve second order equation of motions. That indeed assures ghost instabilities be prevented\footnote{We note that second order equation is not mandatory, see for instance Degenerate Higher Order Scalar-Tensor (DHOST) theories \cite{Langlois:2015cwa,BenAchour:2016fzp}.}, so that just one additional scalar degree $\phi$ is allowed to propagate. Specifically, let us consider the action:
\begin{equation}
    S=\frac{1}{2\kappa}\int d^4x\sqrt{-g}\sum_{i=2}^5 L_i,
\end{equation}
where contributions $L_i$ are given by
\begin{equation}
    \begin{split}
        &L_2=K(\varphi,X)\\
        &L_3=-G_3(\varphi,X) \Box\varphi\\
        &L_4=G_4(\varphi,X)R+G_{4,X}\leri{(\Box\varphi)^2-\Phi_{\mu\nu}\Phi^{\mu\nu}}\\
        &L_5=G_5(\varphi,X)G_{\mu\nu}\Phi^{\mu\nu}+\\
        &\quad+\frac{1}{6} G_{5,X}\leri{(\Box\varphi)^3-3\Box\varphi\,\Phi_{\mu\nu}\Phi^{\mu\nu}+2\Phi\indices{^\mu_\nu}\Phi\indices{^\nu_\rho}\Phi\indices{^\rho_\mu}}.
    \end{split}
\end{equation}
Here $R,\,G_{\mu\nu}$ are the Ricci scalar and the Einstein tensor, respectively, and we introduced the notation
\begin{equation}
    X\equiv-\frac{1}{2}\nabla_\mu\varphi\nabla^\mu\varphi,\quad\Phi_{\mu\nu}\equiv\nabla_\mu\nabla_\nu\varphi,
\end{equation}
We note that by suitable choices of the function $K,G_i$ we can easily reproduce well-knows results, as for instance General Relativity ($G_4=1$ and others vanishing) or metric $f(R)$ theories ($K=-V(\varphi)$ and $G_4=\varphi$). Now, since we want to analyze wave propagation on Minkowski background, we decompose metric as $g_{\mu\nu}=\eta_{\mu\nu}+h_{\mu\nu}$, along with the scalar perturbation $\varphi=\phi_0+\phi$. Then, as explicitly discussed for vacuum in \cite{Hou:2017bqj} (see also \cite{2011PThPh.126..511K} for details), we can rearrange linearized equations in the form
\begin{align}
    &G_{\mu\nu}^{(1)}-\frac{G_{4,\varphi}(0)}{G_4(0)}\leri{\partial_\mu\partial_\nu-\eta_{\mu\nu}\Box}\phi=\kappa''T_{\mu\nu}^{(1)} \label{g wave equations linearized horndeski}\\
    &\leri{\Box-M^2}\phi=\kappa'T^{(1)},
   \label{phi wave equations linearized horndeski}
\end{align}
with the effective mass of the scalar mode given by
\begin{equation}
    M^2=-\frac{K_{,\varphi\varphi}(0)}{K_{,X}(0)-2G_{3,\varphi}(0)+\frac{3G_{4,\varphi}^2(0)}{G_4(0)}},
\end{equation}
and where we also introduced with respect to \cite{Hou:2017bqj} the coupling with matter, described by the effective gravitational constants
\begin{align}
    &\kappa'=\frac{G_{4}(0)\kappa}{G_{4}(0)\leri{K_{,X}(0)-2G_{3,\varphi}(0)}+3G_{4,\varphi}^2(0)}\\
    &\kappa''=\frac{\kappa}{G_{4}(0)},
\end{align}
where the function are all evaluated in $\varphi=\phi_0,\;X=0$. 
Here $T_{\mu\nu}^{(1)},\;T^{(1)}$ represent the perturbations of $\mathcal{O}(h)$ induced on stress energy tensor by $h_{\mu\nu}$ and $\phi$, which as clear from \eqref{g wave equations linearized horndeski} turn out to be dynamically coupled. Now, in order to disentangle truly tensor polarizations from the scalar mode, it is useful to define the generalized trace reversed tensor
\begin{equation}
    \bar{h}_{\mu\nu}\equiv h_{\mu\nu}-\frac{1}{2}\eta_{\mu\nu}\leri{h+\frac{2G_{4,\varphi}(0)}{G_4(0)}\phi},
\end{equation}
with $h=\eta^{\mu\nu}h_{\mu\nu}$. This allows us to rearrange \eqref{g wave equations linearized horndeski} in the well-known form
\begin{equation}
    \Box\bar{h}_{ij}=-2\kappa''T_{ij}^{(1)},
    \label{tensorkleingordon}
\end{equation}
where we imposed transverse and traceless conditions (TT-gauge), i.e. $\partial_\mu\bar{h}^{\mu\nu}=0$ and $\;\bar{h}=0$, which enable us to restrict the analysis to spatial indices. Therefore, even in this extended dynamical framework we can still fully recover standard tensor TT-modes, described by the perturbation $\bar{h}_{\mu\nu}$ and carrying known plus and cross polarizations of General Relativity. Remnant of the coupling with the additional massive mode $\phi$ is then limited to the coupling $\kappa''$, which keeps information about the background value $\phi_0$. In the following, therefore, we will focus on \eqref{tensorkleingordon} and \eqref{phi wave equations linearized horndeski}, setting for the sake of simplicity $\alpha\equiv\frac{G_{4,\varphi}(0)}{G_4(0)}$.

\section{Evaluation of the damping rate}\label{sec2}
\begin{figure}[h]
    \centering
    \includegraphics[width=1\columnwidth]{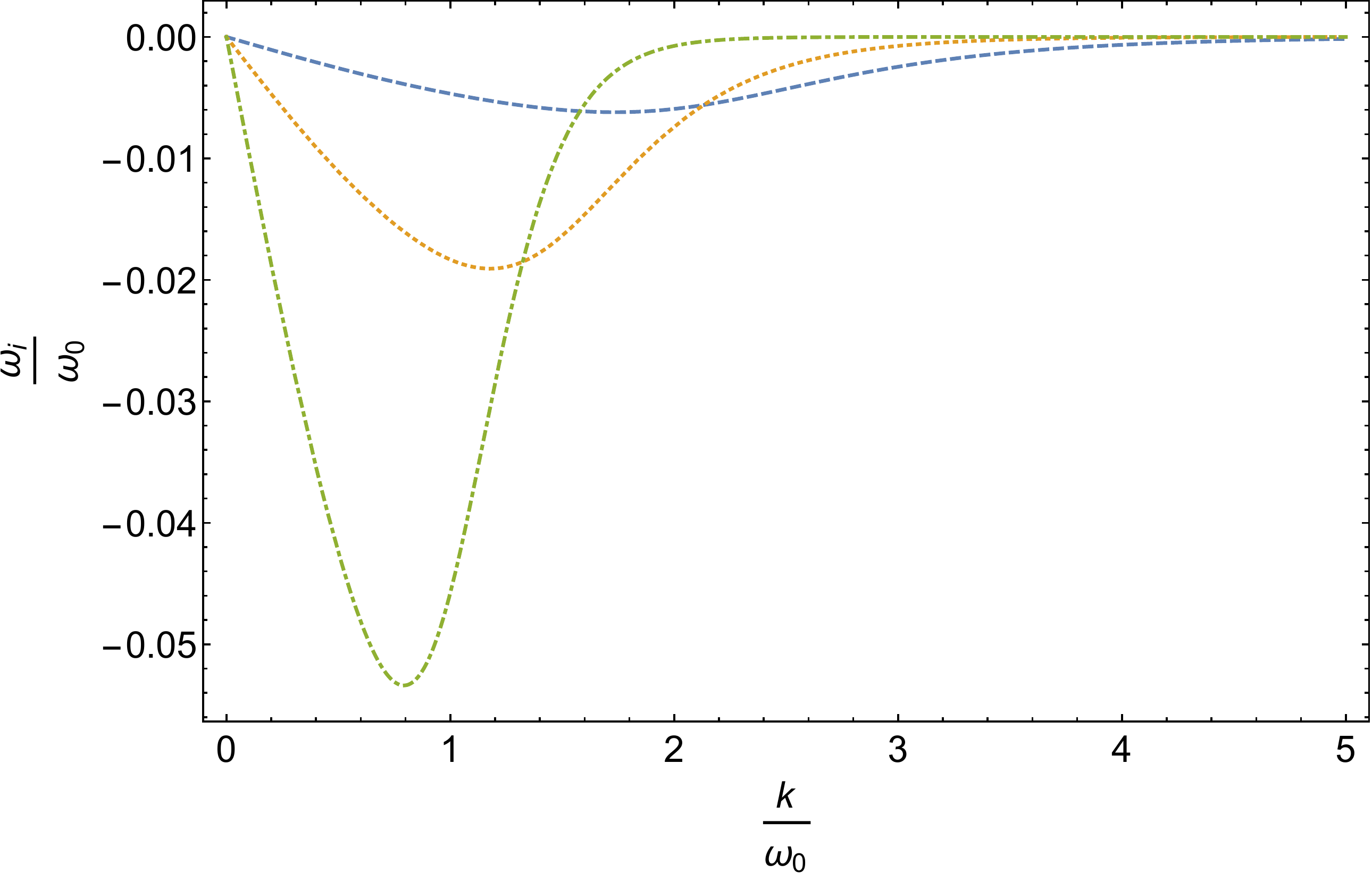}
    \caption{Normalized damping coefficient $\bar{\omega}_i$  vs normalized wavenumber $\bar{k}$ for different values of the parameter $x$: $x=1$ (dashed curve), $x=2$ (dotted curve), $x=5$ (dot-dashed curve). }
    \label{graficheri}
\end{figure}

\begin{figure}[h]
    \centering
    \includegraphics[width=1\columnwidth]{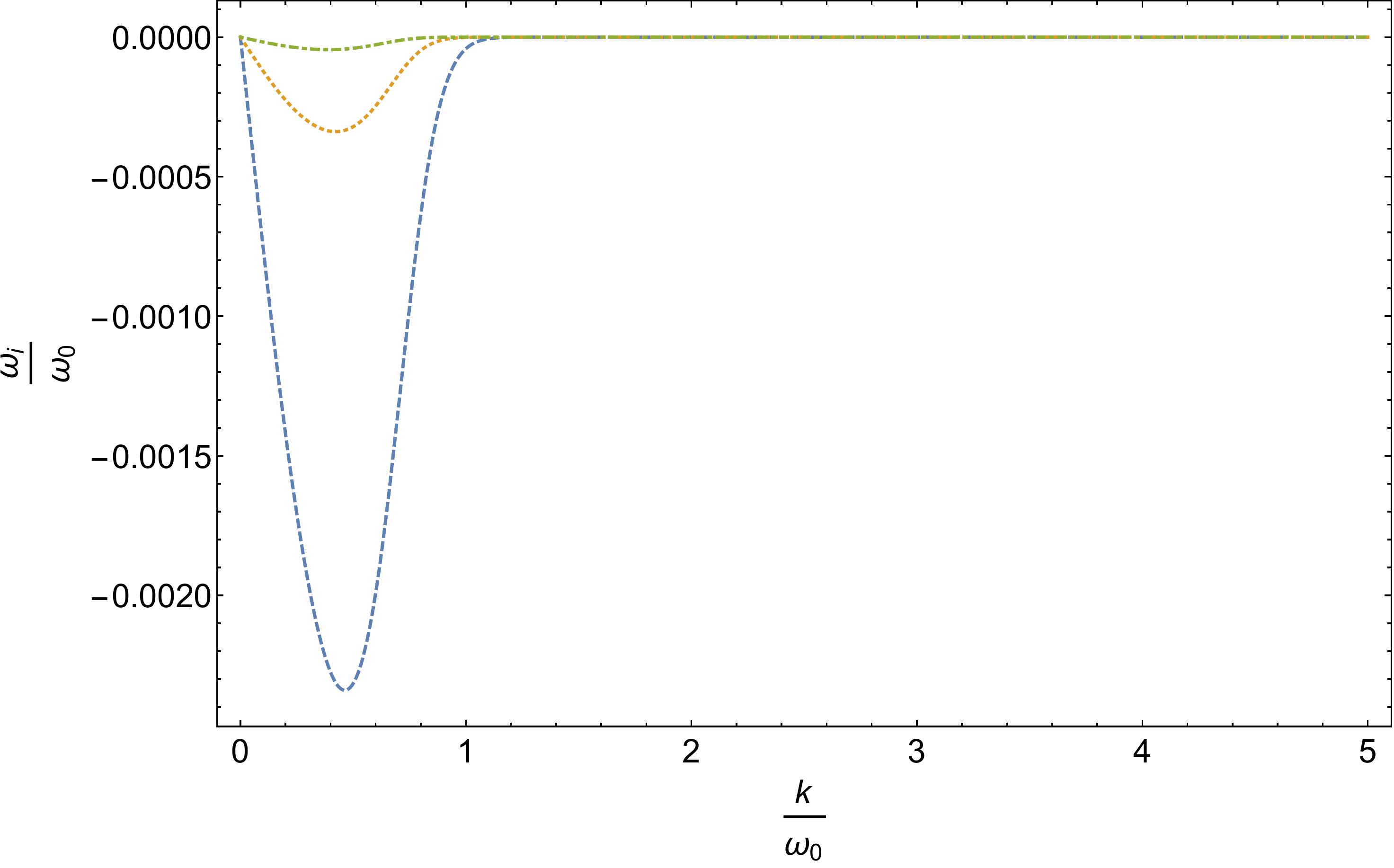}
    \caption{Normalized damping coefficient $\bar{\omega}_i$  vs normalized wavenumber $\bar{k}$ for different values of the parameter $x$: $x=30$ (dashed curve), $x=40$ (dotted curve), $x=50$ (dot-dashed curve). }
    \label{graficheri2}
\end{figure}

 The medium is described by the distribution function $f(\vec{x},\vec{p},t)$, which evolves in time according to the Vlasov equation
\begin{equation}
  \dfrac{D f}{dt}=\dfrac{\partial f}{\partial t}+\dfrac{d x^m}{dt}\dfrac{\partial f}{\partial x^m}+\dfrac{dp_m}{dt}\dfrac{\partial f}{\partial p_m}=0,
\end{equation}
that we write in terms of the covariant spatial components of the momentum $p_m$, following \cite{Flauger:2017ged}. The variation in time of $p_m$ is simply given by
\begin{equation}
    \dfrac{dp_m}{dt}=\dfrac{1}{2p^0}\leri{p_ip_j\dfrac{\partial\bar{h}_{ij}}{\partial x^m}+\alpha\mu^2\dfrac{\partial \phi}{\partial x^m}},
\end{equation}
being $p^0=\sqrt{\mu^2+g^{ij}p_ip_j}$ the particle energy and $\mu$ the particle mass. Then, in the absence of the gravitational wave,  we assume the distribution function be some isotropic equilibrium solution $f_0\leri{p}$ of the unperturbed equation, where $p\equiv\sqrt{\delta^{ij}p_ip_j}$. At the initial time $t=0$ we take $f(\vec{x},\vec{p},0)=f_0( \sqrt{g^{ij}(\vec{x},0)p_ip_j})$ which at the first order results in 
\begin{equation}
   f(\vec{x},\vec{p},0)=f_0 \leri{p}-\dfrac{f_0'(p)}{2}\leri{\dfrac{p_ip_j}{p}\bar{h}_{ij}(\vec{x},0)-\alpha\, p\, \phi(\vec{x},0)},
\end{equation}
where $f_0'(p) \equiv \frac{d f_0}{dp}$.
At any later time, the gravitational wave induces a dynamical perturbation $\delta f (\vec{x},\vec{p},t)$, which we assume to be small with respect to the equilibrium configuration, i.e. $\frac{\delta f}{f_0}=\mathcal{O}(h)$. { Therefore}, in the following we deal with the perturbed distribution function $f(\vec{x},\vec{p},t)+\delta f (\vec{x},\vec{p},t)$. By ignoring all contributes of $\mathcal{O}(h^2)$, we obtain the linearized Vlasov equation for the perturbation $\delta f (\vec{x},\vec{p},t)$:
\begin{equation}
\begin{split}
    &\frac{\partial \delta f }{\partial t}+\frac{p^m}{p^0}\frac{\partial \delta f}{\partial x^m}+\\
    &-\frac{f_0'(p) }{2p} \leri{p_ip_j\frac{\partial \bar{h}_{ij}}{\partial t}-\alpha p^2 \frac{\partial \phi}{\partial t}-\alpha p^0 p^m \frac{\partial \phi}{\partial x^m}}=0.
  \label{linear vlasov}
\end{split}{}
\end{equation}
The dynamical system is now fully determined once the sources of \eqref{phi wave equations linearized horndeski} and \eqref{tensorkleingordon} are evaluated in terms of $f(\vec{x},\vec{p},t)$, namely 
\begin{align}
& T^{(1)}=-\mu^2 \int d^3p\, \dfrac{\delta f (\vec{x},\vec{p},t)}{p^0}\\
\label{traccia}
& T_{ij}^{(1)}=\int d^3p\, \dfrac{p_ip_j}{p^0}\delta f (\vec{x},\vec{p},t),
\end{align}
where $d^3p=dp_1dp_2dp_3$. Then, choosing the $z$ axis to be coincident with the direction of propagation of the gravitational wave, we pursue the analysis in the Fourier-Laplace space. Specifically, we perform a Fourier transform on spatial coordinates accompanied with a Laplace transform on time coordinate $t$. This allows us to solve algebraically the Vlasov equation for the perturbation $\delta f(\vec{x},\vec{p},t)$, i.e.
\begin{widetext}
\begin{equation}
    \delta f^{(k,s)}(\vec{p})=\frac{\frac{f_0'(p)}{2 p}\leri{p_ip_j\leri{s\,\bar{h}_{ij}^{(k,s)}-\bar{h}_{ij}^{(k)}(0)}-\alpha\leri{p^2s+ikp_3p^0}\phi^{(k,s)}+\alpha p^2\phi^{(k)}(0)}}{s+ik \frac{p_3}{p^0}}.
\label{deltafFL}
\end{equation}
\end{widetext}{}
where the Fourier and Fourier-Laplace components of the fields are displayed as $\phi^{(k)}(t)$ and $\phi^{(k,s)}$, respectively\footnote{Analogously for the fields $\bar{h}_{ij}(z,t)$ and $\delta f(z,\vec{p},t)$. }. Similarly, considering the Fourier-Laplace transform for \eqref{phi wave equations linearized horndeski} and \eqref{tensorkleingordon} we get\footnote{Without loss of generality, we set the initial conditions $\dot{\phi}^{(k)}(0)$, $\dot{\bar{h}}_{ij}^{(k)}(0)$ and $\delta f^{(k)}(0)$ vanishing.
}\\
\begin{align}
    &\phi^{(k,s)}=\frac{s\phi^{(k)}(0)+\mu^2\kappa'\int d^3p \frac{\delta f^{(k,s)} (\vec{p})}{p^0}}{s^2+k^2+m^2}
    \label{FLkleingordon}\\
    &\bar{h}_{ij}^{(k,s)}=\frac{s\bar{h}_{ij}^{(k)}(0)+2\kappa''\int d^3p \frac{p_ip_j}{p^0}\delta f^{(k,s)} (\vec{p})}{s^2+k^2}.
    \label{FLtensorkleingordon}
\end{align}
Now, even if $\delta f^{(k,s)}(\vec{p})$ actually depends both on $\phi^{(k,s)}$ and { on} $\bar{h}_{ij}^{(k,s)}$, calculations show that the solutions of \eqref{FLkleingordon} and \eqref{FLtensorkleingordon} are indeed uncoupled, i.e.
\begin{align}
    &\phi^{(k,s)}=\dfrac{\leri{s+\alpha\pi \mu^2 \kappa'\int d\rho dp_3\, \rho\, \dfrac{f_0'(p) p }{p^0s+ikp_3}}\phi^{(k)}(0)}{(s^2+k^2+m^2)\epsilon^{(\phi)}(k,s)}\label{equationphiuncoupled}\\
    &\bar{h}_{ij}^{(k,s)}=\dfrac{\leri{s-\frac{\pi \kappa''}{2}\int d\rho dp_3 \,\rho^5 \dfrac{f'_0(p)}{p(p^0s+ikp_3)}}\bar{h}_{ij}^{(k)}(0)}{(s^2+k^2)\epsilon^{(h)}(k,s)},
    \label{equationhuncoupled}
\end{align}
where integrals have been conveniently restated in cylindrical coordinates by performing the substitution $\leri{p_1,p_2}\rightarrow \leri{\rho,\chi}$, with $\rho^2=p_1^2+p_2^2$ and $\tan(\chi)=\frac{p_2}{p_1}$.
By analogy with the electromagnetic theory we introduced the complex dielectric functions
\begin{equation}
    \label{epsilonscalar}
\epsilon^{(\phi)}(k,s)=1
    +\dfrac{\alpha \pi\mu^2 \kappa'}{s^2+k^2+m^2}\int d\rho dp_3\,\rho \dfrac{f_0'(p)}{p} \dfrac{p^2s+ikp^0p_3}{p^0s+ikp_3}
\end{equation}
\begin{equation}
    \label{epsilontensor}
\epsilon^{(h)}(k,s)=1
    -\dfrac{\pi\kappa''}{2(s^2+k^2)}\int d\rho dp_3\,\rho^5 \dfrac{f_0'(p)}{p} \dfrac{s}{p^0s+ikp_3},
\end{equation}{}
and the settling of damping can be inferred by the behaviour of the real and imaginary part\footnote{In the following real and imaginary part are denoted by subscript $r$ and $i$, respectively.} of $\epsilon^{(\phi,h)}(k,\omega)\equiv\epsilon^{(\phi,h)}(k,s=-i\omega)$, where we defined the complex frequency $\omega= \omega_r + i \omega_i$. In particular, when the condition $|\omega_r| \gg |\omega_i|$ holds (i.e. weak damping scenario \cite{Landau:1946jc,pitaevskii2012physical}), the oscillation period is much smaller than the damping time, and we properly deal with a damped wave rather than a transient. In this case, thus, the dispersion relation $\omega_r=\omega_r(k)$ can be derived by solving
\begin{equation}
    \epsilon^{(\phi,h)}_r(k,\omega_r)=0\label{realpart},
\end{equation}
while the damping coefficient is obtained from
\begin{equation}
    \quad\omega_i=-\left.\frac{ \epsilon^{(\phi,h)}_i(k,\omega)}{\frac{\partial \epsilon^{(\phi,h)}_r(k,\omega)}{\partial\omega}}\right|_{\omega=\omega_r}.
    \label{imaginarypart}
\end{equation}\\
We assume that the equilibrium distribution function is the J\"{u}ttner-Maxwell distribution
\begin{equation}
    f_0(p)=\dfrac{n}{4 \pi \mu^2 \Theta K_2 \leri{\frac{\mu}{\Theta}}}e^{-\frac{\sqrt{\mu^2+p^2}}{\Theta}}
\end{equation}
where $n$ is the density of particles, $\Theta$ is the temperature in units of the Boltzmann constant $k_B$ and $K_2\leri{\cdot}$ is the modified Bessel function of the second kind. Then, in the presence of such a distribution { function}, the dielectric functions \eqref{epsilonscalar}-\eqref{epsilontensor} assume the form:
\begin{widetext}
\begin{align}
    &\epsilon^{(\phi)} (k,\omega_r)=1-\frac{\alpha\kappa' n }{2 \leri{k^2+m^2-\omega_r^2}\Theta^2K_2\leri{\frac{\mu}{\Theta}}}\int_0^{+\infty} d\rho \,\int _{0}^{+\infty} dp_3 \, \dfrac{\rho\leri{p_3^2-\frac{v_p^2 }{1-v_p^2}\rho^2}}{p_3^2-\frac{v_p^2 }{1-v_p^2}(\mu^2+\rho^2)}\,e^{-\frac{\sqrt{\mu^2+\rho^2+p_3^2}}{\Theta}}
    \label{dielectriccompletephi}\\
    &\epsilon^{(h)} (k,\omega_r)=1-\frac{\kappa'' n}{4 \leri{k^2-\omega_r^2}\mu^2\Theta^2K_2\leri{\frac{\mu}{\Theta}}}\dfrac{v_p^2}{1-v_p^2}\int_0^{+\infty} d\rho \, \int _{0}^{+\infty} dp_3 \, \dfrac{\rho^5}{p_3^2-\frac{v_p^2 }{1-v_p^2}(\mu^2+\rho^2)}\,e^{-\frac{\sqrt{\mu^2+\rho^2+p_3^2}}{\Theta}} \label{dielectriccompleteh}
\end{align}
\end{widetext}
where $v_p\equiv\frac{\omega_r}{k}$ is the phase velocity. Now, by virtue of the condition $|\omega_r|\gg |\omega_i|$ and provided $v_p<1$, integrals in \eqref{dielectriccompletephi}-\eqref{dielectriccompleteh} are featured by a pair of poles on the real axis, located at the points $p_3=\pm\sqrt{\frac{v_p^2}{1-v_p^2}(\mu^2+\rho^2)}$. That guarantees the existence of the imaginary part for dielectric functions, stemming from the integration around poles and evaluable with the residue theorem, ultimately responsible for the appearing of the damping coefficient \eqref{imaginarypart}. We note, however, that condition $v_p< 1$ is not a priori ensured and its attainability has to be indeed verified by calculating $\omega_r$ from \eqref{realpart}. As usually done in plasma physics \cite{pitaevskii2012physical}, we calculate the Langmuir dispersion relation by expanding the denominator contained in the integrals up to the second order in $p_3$, under the assumption $\frac{p_3\sqrt{1-v_p^2}}{v_p\sqrt{\mu^2+\rho^2}}\ll 1$, which corresponds to the requirement of having a phase velocity for the wave much greater than the thermal velocity of the medium. 
\\ Then, explicit calculations for tensor modes lead to 
\begin{equation}
    \epsilon^{(h)}_r(k,\omega_r)=1+\frac{12v_{h}^2}{x^2}\leri{\frac{x}{1-v_p^2}+\frac{\gamma(x)}{v_p^2} }=0,
    \label{realpart2}
\end{equation}{}
where we introduced the dimensionless quantity $x\equiv\frac{\mu}{\Theta}$, in terms of which  $\gamma(x)\equiv\frac{K_1(x)}{K_2(x)}$. Here, $v_h^2=\frac{\omega_h^2}{k^2}\equiv\frac{\kappa'' n \mu}{6k^2}$ can be seen as the proper phase velocity of the medium for tensor excitations (see \cite{Montani:2018iqd} for a comparison). Now, the solution of \eqref{realpart2} is given by
\begin{equation}\label{disp1}
    v_p^2=\frac{1+12v_h^2\frac{x-\gamma(x)}{x^2}+\sqrt{\leri{1+12v_h^2\frac{x-\gamma(x)}{x^2}}^2+48v_h^2\frac{\gamma(x)}{x^2}}}{2},
\end{equation}{}
which can be easily demonstrated to be identically greater than unity, { implying that tensor modes cannot be damped when travelling throughout the medium. We stress that is in agreement both with} their purely transverse nature, and with the absence of coupling with the scalar mode in linearized equations \eqref{equationphiuncoupled}-\eqref{equationhuncoupled}. Therefore, in the following we will restrict our analysis to scalar perturbations. In this case the dielectric function real part turns out to be
\begin{equation}
    \epsilon_r(k,\omega)= 1+3\gamma(x)\omega_0^2\dfrac{1-3v_p^2}{v_p^2(k^2+m^2-\omega^2)},    \label{epsilonphireale}
\end{equation}
with $\omega_0^2$ defined by analogy with the tensor case, i.e. $\omega_0^2\equiv\frac{\alpha \kappa' n \mu}{6}$, and $v_p$ referring solely to the scalar polarization. Hence, solving \eqref{realpart} for \eqref{epsilonphireale}, we obtain 
\begin{equation}
       \omega_r^2=\dfrac{k^2+m^2-9\gamma\omega_0^2 +\sqrt{\leri{k^2+m^2-9\gamma\omega_0^2}^2+12\gamma\omega_0^2 k^2}}{2},
   \label{realpartomega}
\end{equation}
and in order to have $v_p<1$ the following constraint must hold
\begin{equation}
    m^2<6\gamma\omega_0^2.
    \label{conditionfordamping}
\end{equation}
This condition, by relating model depending parameters with physical quantities describing the medium, allows to select theories of gravity sensitive to damping from non-interacting ones. 
\\ It has to be noticed that dispersion relations \eqref{disp1}-\eqref{realpartomega} imply superluminal group velocities in specific regions of $k$. In many contexts group velocity is associated with energy and information transport, therefore a superluminal behavior would imply causality violation, but this is not our case: we recall that we have perturbed the medium with a wave which is non-null at the initial time $t=0$ but identically vanishes for any negative time. The discontinuity at the initial time causes the emergence of a front wave, namely a high frequency Sommerfeld precursor \cite{sommerfeld}, propagating at the speed of light. It is a well established result \cite{brillouin} that, in such cases, energy and information transport is associated to the front wave propagation.
\\ Then, we calculate by means of the residue theorem the imaginary part of the dielectric function, resulting in
\begin{equation}\label{epsiloni}
    \epsilon_i(k,\omega_r)= -\dfrac{3\pi \omega_0^2 x}{2K_2(x)}\frac{v_pe^{-\frac{x}{\sqrt{1-v_p^2}}}}{k^2+m^2-\omega_r^2},
\end{equation}
which inserted in \eqref{imaginarypart} gives us finally $\omega_i$, that is
\begin{equation}
    \omega_i=-\frac{\pi x\,k}{4 K_1(x) }\frac{v_p^4(1+\frac{m^2}{k^2}-v^2_p)e^{-\frac{x}{\sqrt{1-v_p^2}}}}{3v^4_p-2v^2_p+(1+\frac{m^2}{k^2})}.
\end{equation}{}
\label{immaginarypartomega}
We emphasize that by virtue of \eqref{realpartomega} the condition $\omega_i<0$ identically holds, and the theory is devoid of instabilities attributable to enhancement phenomena. 
It must be stressed that the analysis { is} pursued under the hypothesis { that the} phase velocity of the wave { is} much greater than the thermal velocity of particles, { so that it} does not apply satisfactorily when highly relativistic media are considered. Indeed, in the cases in which the parameter $x$ approaches the unity or smaller values, the particles are characterized by a thermal velocity almost equal to the speed of light, therefore the aforementioned ansatz is not viable. In order to have predictive power also in these scenarios we treat the problem with numerical techniques. In particular we calculate \eqref{dielectriccompletephi} through a numerical integration, assuming that the result of the latter corresponds to the real part of the dielectric function, whilst we retain the imaginary part \eqref{epsiloni}, exactly calculated as a residue around the pole. The analysis { is} carried out under the ulterior hypothesis that the quantity $\delta=1-v_p$ be positive and much smaller than unity. We obtain a damping coefficient $\bar{\omega}_i=\bar{\omega}_i \leri{\bar{k}; x,\bar{m}}$, where barred quantities are intended as normalized with the proper frequency $\omega_0$, e.g. $\bar{\omega}_i\equiv \frac{\omega_i}{\omega_0}$ and equivalently for the others \footnote{In the following we will set, for the sake of simplicity, $\alpha=1$ and $\kappa'=\kappa$ in the expression of the proper frequency.}. { It shows} remarkable analogies with the damping coefficient calculated with the analytic treatment, as for instance to be negative for any value of the wavenumber and { to} be identically null when $\bar{m}$ exceeds the bound \eqref{conditionfordamping}. We report in FIG. \ref{graficheri} the curves obtained for three different values of the parameter $x$, keeping the normalized mass $\bar{m}$ fixed to 1.
A qualitative analysis shows that the minimum, i.e. the maximum damping, is localized at $\bar{k} \simeq \bar{m}$. Moreover, the effect is rapidly suppressed if relativistic conditions are not fulfilled, i.e. when the parameter $x$ becomes larger than few tens, as clearly observable in FIG. \ref{graficheri2}.
The need to look at relativistic media, characterized by a high density, suggests the application of our formulae to a cosmological scenario. This will be the objective pursued in the subsequent section.  

\section{Cosmological implementation}\label{sec3}
\begin{figure}[h]
    \centering
    \includegraphics[width=1\columnwidth]{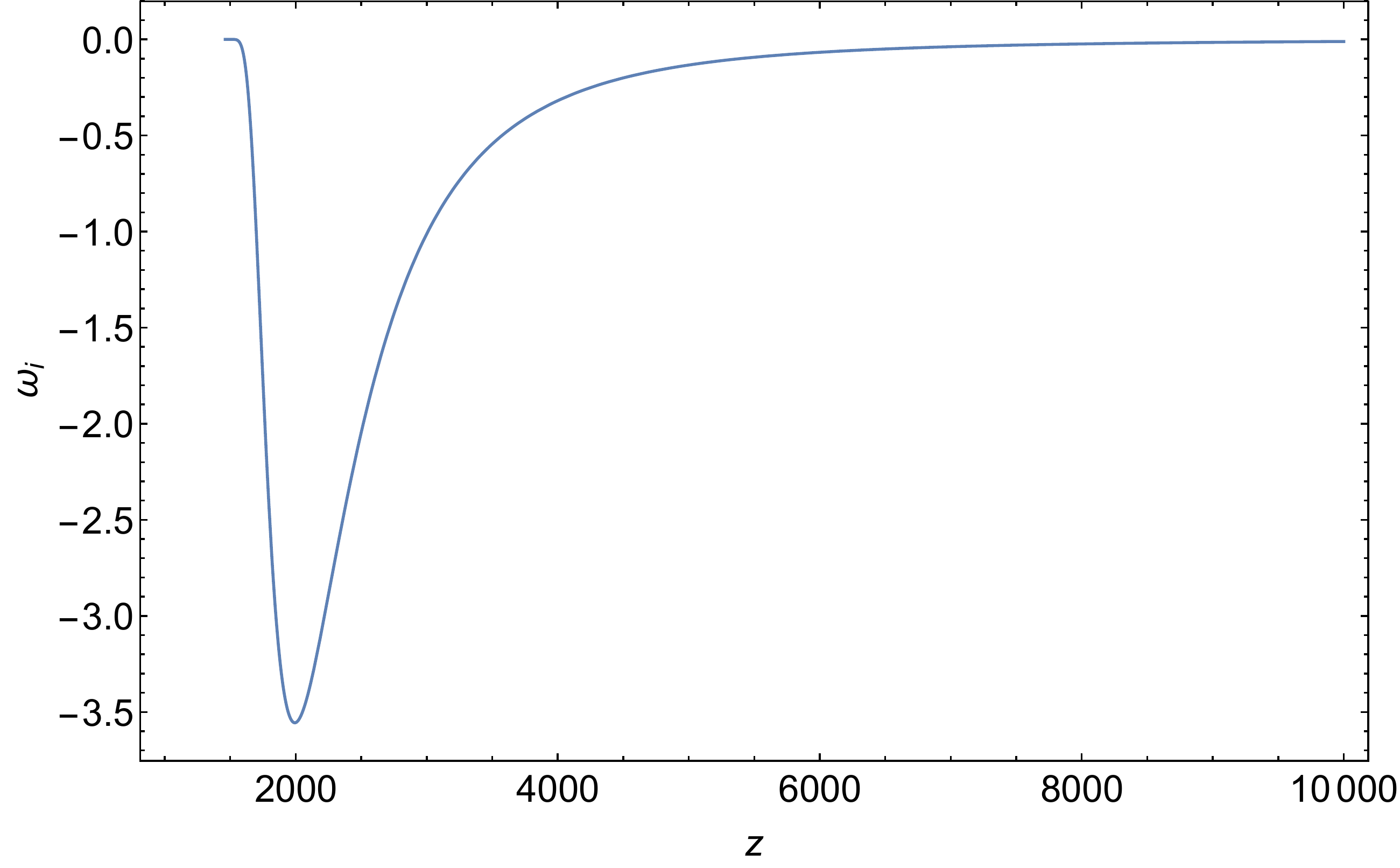}
    \caption{Damping coefficient $\omega_i$ (in $10^{-16}$ Hz) vs redshift $z$. }
    \label{gammadiz}
\end{figure}

\begin{figure}[h]
    \centering
    \includegraphics[width=1\columnwidth]{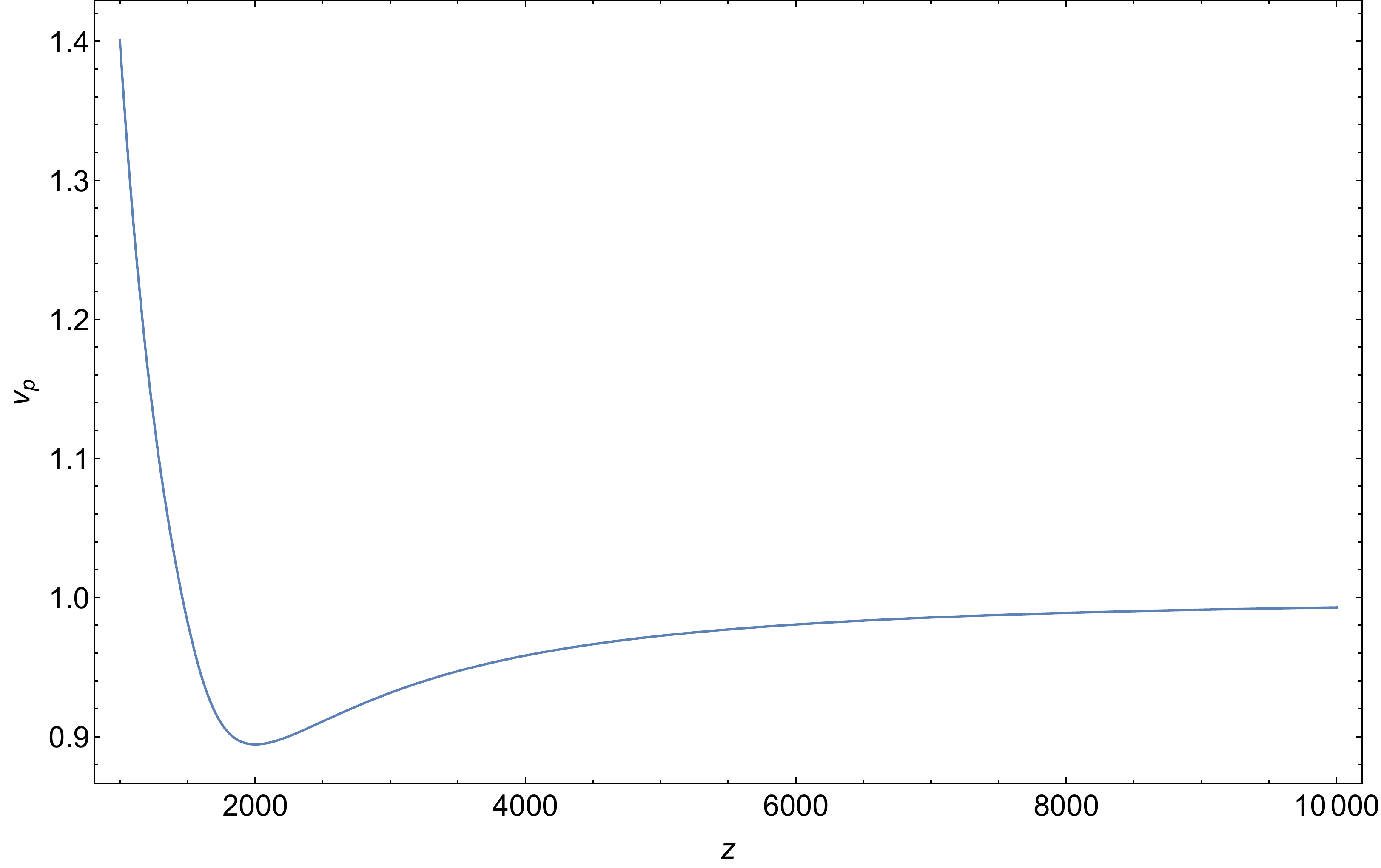}
    \caption{Phase velocity $v_p$ vs redshift $z$. }
    \label{vdifase}
\end{figure}

Before { we apply} our findings to a cosmological setting, some explanations about the feasibility of such { an} implementation are needed. It is worth noting that our model must be intended as referred to a local inertial frame, where Vlasov equation is properly defined and where the curvature of the background can be neglected. In a practical sense, such an hypothesis corresponds to consider gravitational waves having a wavelength much smaller than the background curvature. { Locally}, the medium appears enough homogeneous and isotropic, and the longitudinal massive mode exhibits a non-zero decaying rate, i.e. Landau damping settles down.
A typical effect of the Universe expansion on gravitational waves consists in the redshift of their amplitude, as a consequence of the time scaling of each spatial length, including their wavelength. More specifically, { the rate of expansion} acts as a friction term on the gravitational wave propagation \cite{Weinberg:2008zzc,Montani:2011zz}. Additional damping effects, due to the neutrino anisotropic stress and associated to the electro-weak transition phase, has been discussed in \cite{Weinberg:2003ur,Lattanzi:2005xb}. Clearly, the spatial curvature of a cosmological background is essentially negligible and our hypothesis of local homogeneity and isotropy for the medium { can be} very well satisfied. Nonetheless, the spacetime curvature is still present via the non stationary character of the cosmological space. This feature could appear in contrast with the physical scheme above, since no expanding background or non-stationary effect are included in the Vlasov equation. However, it is a well-known fact that any physical process which takes place on a physical scale much smaller than the Hubble micro-physical scale $L_H\sim H^{-1}$ ($H$ being the Hubble function), it is not significantly affected by the Universe expansion. This is just the reason because well-inside the scale $L_H$, the dynamics of the cosmological density fluctuations can be treated on a Newtonian level, leading to the concept of Jeans length \cite{Weinberg:2008zzc,Montani:2011zz}. The same situation holds for the tensor perturbations, associated to cosmological gravitational waves. Actually, the physical condition to be satisfied in order the Universe expansion rate could be neglected in the wave propagation, corresponds to require that the physical wavelength of the propagating mode $\lambda _{phys}$, i.e. scaled for the cosmic scale factor, be much smaller than the Hubble scale $L_{H}$.
On a cosmological level, this requires the spacetime curvature be much greater than the { propagating} wavelength. By other words, the cosmological background gravity must be much smaller than the gravitational field due to the waves themselves. At this level, therefore, the so-called Jeans swindle is overcome because the spatial distribution of matter is really homogeneous and infinite. That is to say, the Newtonian gravitational field within the Hubble scale can be properly considered as a vanishing contribution, as Jeans assumed in its original formulation (the Jeans swindle has a value when the matter distribution is homogeneous, but with a finite extension). However, it is worth noting that Vlasov equation does not apply to the cosmological medium, for which the collision term is far from being zero, since the collisions must preserve the equilibrium of the thermal bath. Thus, our analysis must be applied to species which are already decoupled from the cosmological thermal bath, and having a sufficient temporal range of interaction with longitudinal massive modes in order its effect on the amplitude attenuation be appreciated.
{ Taking} the Cosmic Microwave Background Radiation (CMBR) as the observational { setting for seeking} effects of such a longitudinal mode, the maximum temporal range of interaction of a decoupled { species} with the waves is the recombination age. It is rather immediate to recognize that the most natural decoupled component of the Universe to be considered is { then} the so-called \emph{cold dark matter}, { composed of} weakly interacting { particles} which decouple from the thermal bath in
the very early Universe.
\\Before applying our formulae to a defined dark matter model, it has to be outlined that in the cosmological implementation the quantities involved acquire a dependence from the redshift $z$, i.e.
 \begin{equation}
     \begin{split}
         k&=k^{(0)} (1+z) \\
         \omega_0&=\omega_0^{(0)} (1+z)^{\frac{3}{2}}\\
         x&=x^{(0)}(1+z)^{-1},
     \end{split}
 \end{equation}
 where we indicate with the superscript $(0)$ present-day values. Hence, by fixing the masses $m$ and $\mu$ together with the current density $n^{(0)}$, it is possible to trace back in time the behavior of the damping coefficient as a function of the wavenumber. As previously stated, the maximum damping occurs around  $\bar{k}\simeq \bar{m}$, therefore it is possible to set $k^{(0)}$ in order to have $\bar{k}=\bar{m}$ at some privileged $z_0$ and maximize the effect at that redshift. As a result, we obtain an expression for the damping coefficient which solely depends on $z$. Further, the wavenumber must obey the bound $k L_H >1$ during the entire time of interaction, together with the constraint of being observable with recent CMBR measurements, which roughly results in $k^{(0)}\lesssim 0.5 \, \text{Mpc}^{-1}$ \cite{Aghanim:2018eyx}. The request to look at sub-horizon modes is more easily satisfied by considering redshifts which are as close as possible to the recombination. Hence, we analyze the case of an interaction ending right before $z\simeq 1100$, with its peak around $z \simeq 2000$. We choose $k^{(0)}=5.6\cdot 10^{-3} \, \text{Mpc}^{-1}$ and $m=2.5\cdot 10^{-28}\, \text{eV}$, so that $\bar{k}=\bar{m}$ is realized at $z\simeq 2000$, while keeping $m$ below the current constraints on the graviton mass \cite{Abbott:2017vtc}. Moreover, we fix $\mu=1.75 \; \text{eV}$ in order to have $x \approx 5 $ at the same redshift and $n^{(0)}$ by imposing the present-day value of dark matter density, i.e. $\mu n^{(0)}=10^{-27}\, \frac{\text{kg}}{\text{m}^3}$.
    We report the plot of the damping coefficient as function of the redshift for the chosen values of the physical parameters in FIG. \ref{gammadiz}.
    It can be observed that the effect is negligible when $z\gtrsim 10^4$, whereas it is strictly null if $z \lesssim 1500$.
 The cause of this latter fact is that, for decreasing redshifts, the normalized mass $\bar{m}$ grows { more} rapidly than $x$. Therefore, for any defined model of massive medium traversed, there exists a redshift below which inequality \eqref{conditionfordamping}
does not hold anymore, causing the phase speed be superluminal and { the} vanishing of { the} damping effect. This can be further appreciated by evaluating the phase velocity as a function of the redshift, as done for the damping coefficient. We report the plot obtained in FIG. \ref{vdifase}.
The expression of the damping coefficient as { a} function of { the} redshift is also useful to obtain the magnitude of the time-integrated effect. If we normalize the amplitude of the wave to unity at some time $t_0$, then the amplitude at any time $t >t_0$ is given by
\begin{equation}\label{attenuazione}
    \mathcal{A}(t)=e^{\int_{t_0}^t \omega_i(t')\, dt'}.
\end{equation}
If we consider $z>3400$, i.e. we look at the radiation-dominated era, the relation between the coordinate time and the redshift is
\begin{equation}
    z(t) \simeq \leri{2 H_0 \sqrt{\Omega_{0,r}} t}^{-\frac{1}{2}},
\end{equation}
with $H_0$ and $\Omega_{0,r}$ the present-day values \footnote{The values of the
cosmological parameters involved in our analysis are taken from \cite{Aghanim:2018eyx}.} of Hubble constant and radiation density, respectively, which we set as $H_0=2.19\; \text{aHz}$ and $\Omega_{0,r}=9.2\cdot 10^{-5}$. During the matter-dominated expansion, namely { for} $z<3400$, we can instead assume 
\begin{equation}
    z(t) \simeq \leri{\dfrac{3}{2} H_0 \sqrt{\Omega_{0,m}} t}^{-\frac{2}{3}},
\end{equation}
where in this case $\Omega_{0,m}=0.315$ is the present-day value of matter density.
By performing the change of variables $t \to z$ in \eqref{attenuazione}, we integrate from $z_0=10^4$ to $z=1450$. The departure from unity of the amplitude, i.e. the absorption, turns out to be in this case
\begin{equation}\label{relabs}
 1-\mathcal{A}(z)=8.2\cdot 10^{-4}.
\end{equation}
The value of relative absorption that we obtain with this simple model is small enough to claim that Landau damping does not significantly affect the observability of longitudinal scalar modes on the CMBR perturbation spectrum. However, it must be pointed out that the absorption we calculated is pertaining to a specific value of the wavenumber: by considering a different $k^{(0)}$, sufficiently close to { that} one we chose in our example, the absorption { rate would} acquire a value close to that reported in \eqref{relabs}. { Conversely}, by looking to a ten-times larger wave number $k^{(0)}=5 \cdot 10^{-2}\, \text{Mpc}^{-1}$, one would obtain an extremely small relative absorption of order $10^{-30}$. Hence, a specific model of massive medium acts as a filter on a narrow ranges of wavelengths, depending on the chosen values of the physical parameters which characterize it. The implications of this peculiar feature are to be taken into account when analyses of scalar perturbations are performed in the spirit of highlighting departures from General Relativity.

\section{Concluding remarks}\label{sec4}
Our analysis clearly establishes a new point of view on the interaction of gravitational waves with astrophysical media: longitudinal scalar modes can be damped by non collisional ensembles of massive particles. In particular, even if analytic estimations can be pursued only in the specif regime of low relativistic media, i.e. { for gravitational waves with phase velocity greater than the thermal velocity of particles}, numerical studies can be easily extended to deal with properly relativistic frameworks.
The application to { a simplified} cosmological settings shows that, at the physical scales for which our treatment holds valid, Landau damping effect is unable to significantly suppress cosmological longitudinal modes. { In principle, that should not prevent the observability of such additional scalar modes on the CMBR radiation, at least for the range of wavenumbers taken into account.} Despite this, we highlighted the possibility of a relative absorption of order $10^{-3}$, if a medium compatible with viable dark matter models is considered at a redshift between $10^3$ and $10^4$. { This effect, in particular,} is associated to wavelengths satisfying the condition of being sub-horizon during the entire time of interaction, and inside the sensitivity curves of recent CMBR observations. We claim, { therefore,} that the damping { due to the interaction} with the decoupled dark matter medium can act as a filter on specific wavelengths, giving rise to a possible modification of the original power spectrum. { We remark, however, that our treatment does not apply in the presence of the cosmological bath, where collisions must be properly included.} 
{ Moreover, the situation can} be significantly different if larger { longitudinal modes} are considered. In the case of Hubble sized scalar fluctuations, { indeed,} the present analysis must be deeply revised since the non-stationary character of the cosmological background can be no longer neglected. { We refer, in particular, to the need of quantifying the relative amount of the Landau damping with respect to the damping coming from the cosmological redshift.} 
\\{ To conclude, we note also that our analysis relies on a perturbative approach, and the inclusion of higher order effects can significantly alter the outcomes. We refer, especially, to} the settling of quasi-linear interactions between longitudinal gravitational waves and massive media \cite{briggs,drummond1962nucl,doi:10.1063/1.1692190}, { which in plasma lead} to appreciable modifications of the medium distribution function, for velocities which are comparable with the phase velocity of the wave.
\section*{Acknowledgements}
\noindent The authors thank Nakia Carlevaro for the fruitful discussions.
\section*{Funding}
\noindent The work of F. B. is supported by the Fondazione Angelo della Riccia grant for the year 2020.
\bibliographystyle{unsrt}
\bibliography{LDamping}
\end{document}